%% file: SMVDI.tex
\documentclass[12pt, a4paper, twoside]{article}

\usepackage{fancyhdr}
\usepackage[utf8]{inputenc}
\usepackage[american]{babel}	
\usepackage{natbib}

\usepackage{fix-cm}
\usepackage{mathtools}
\usepackage{amsfonts}
\usepackage{dsfont}
\usepackage{bm}	

\usepackage{graphicx}
\usepackage{tikz}
\usetikzlibrary{shapes.geometric, arrows}
\tikzstyle{start} = [rectangle, rounded corners, minimum width=3.25cm, text width=3.25cm, minimum height=1cm, text centered, draw=black, fill=red!30]
\tikzstyle{stop} = [rectangle, rounded corners, minimum width=3.25cm, text width=3.25cm, minimum height=0.5cm, text centered, draw=black, fill=green!30]
\tikzstyle{decision} = [rectangle, rounded corners, minimum width=3.25cm, text width=3.25cm, minimum height=1cm, text centered, draw=black, fill=blue!30]
\tikzstyle{process} = [rectangle, rounded corners, minimum width=3.25cm, minimum height=1cm, text centered, text width=3.25cm, draw=black, fill=orange!30]
\tikzstyle{io} = [diamond, minimum width=3.25cm, text width=3.25cm, minimum height=1cm, text centered, draw=black, fill=green!30]
\tikzstyle{arrow} = [thick,->,>=stealth]
\tikzstyle{every node} = [font=\small]

\usepackage{enumitem}

\usepackage{url} 

\allowdisplaybreaks	

\newcommand{\beqo}{\begin{equation*}}
\newcommand{\eeqo}{\end{equation*}}
\newcommand{\beqq}{\begin{equation}}
\newcommand{\eeqq}{\end{equation}}

\newcommand{\N}{\mathbb{N}}
\newcommand{\Z}{\mathbb{Z}}
\newcommand{\R}{\mathbb{R}}

\newcommand{\calK}{\mathcal{K}}
\newcommand{\calA}{\mathcal{A}}

\newcommand{\calN}{\mathcal{N}}
\newcommand{\calM}{\mathcal{M}}

\newcommand{\veps}{{\varepsilon}}
\newcommand{\eps}{{\epsilon}}

\newcommand{\te}{{\theta}}

\newcommand{\si}{{\sigma}}

\newcommand{\btheta}{\bm \theta}

\newcommand{\beins}{{\bm 1}}

\newcommand{\bu}{{\bm{u}}}

\newcommand{\bv}{{\bm{v}}}
\newcommand{\bw}{{\bm{w}}}
\newcommand{\bX}{{\bm{X}}}
\newcommand{\bx}{{\bm{x}}}

\newcommand{\bz}{{\bm{z}}}

\DeclareMathOperator{\rank}{rank}

\DeclareMathOperator{\SI}{SI}
\DeclareMathOperator{\SPI}{SPI}
\DeclareMathOperator{\SPEI}{SPEI}
\newcommand{\ASMI}{\text{SMI}^\calA}
\newcommand{\MSMI}{\text{SMI}^\calM}
\newcommand{\NSMI}{\text{SMI}^\calN}

\newcommand{\wh}{\widehat}
\newcommand{\wt}{\widetilde}

\newcommand{\Ind}{{\mathds 1}}
\newcommand{\ind}[1]{\Ind{\left\{#1\right\}}}

\newcommand{\T}{'}	

\begin{document}

\title{Standardized drought indices:\\ A novel uni- and multivariate approach}

\author{by\\
\\
Tobias M. Erhardt\thanks{TUM International Graduate School of Science and Engineering (IGSSE)} ~and Claudia Czado\\
Zentrum Mathematik\\
Technische Universit\"at M\"unchen\\
Boltzmannstr. 3, 85748 Garching, Germany}

\date{\today}
\maketitle

\noindent\rule{\textwidth}{0.1pt}\\
	
\section*{Abstract}
	As drought is among the natural hazards which affects people and economies worldwide and often results in huge monetary losses sophisticated methods for drought monitoring and decision making are needed. Several different approaches to quantify drought have been developed during past decades. However, most of these drought indices suffer from different shortcomings and do not account for the multiple driving factors which promote drought conditions and their inter-dependencies. We provide a novel methodology for the calculation of (multivariate) drought indices, which combines the advantages of existing approaches and omits their disadvantages. Moreover, our approach benefits from the flexibility of vine copulas in modeling multivariate non-Gaussian inter-variable dependence structures. A three-variate data example is used in order to investigate drought conditions in Europe and to illustrate and reason the different modeling steps. The data analysis shows the appropriateness of the described methodology. Comparison to well-established drought indices shows the benefits of our multivariate approach. The validity of the new methodology is verified by comparing the spatial extent of historic drought events based on different drought indices. Further, we show that the assumption of non-Gaussian dependence structures is well-grounded in this real-world application.
	
\noindent\textit{Keywords:} standardized drought indices, dependence modeling, drought modeling, vine copulas
	
\noindent\rule{\textwidth}{0.1pt}

\fancyhf{}
\fancyhead[OR]{\textit{Standardized drought indices: A novel uni- and multivariate approach} \qquad\thepage}
\fancyhead[EL]{\thepage\qquad \textit{T. M. Erhardt and C. Czado}}
\renewcommand{\headrulewidth}{0pt}

\input{Sections/Introduction.tex}

\input{Sections/Data.tex}

\input{Sections/Univariate.tex}

\input{Sections/Multivariate.tex}

\input{Sections/Verification.tex}

\input{Sections/Conclusions.tex}

\section{Supporting information}

\subsection{Accompanying figures and analyses}\label{suppA}

We provide figures and further analyses of the data at hand, visualizing/complementing the presented methodology for drought index calculation. We address the following issues:
\begin{enumerate}
	\item Visualization of the data and its features
	\item Testing of multivariate normality
	\item Visualization of the area affected by drought according to the different indices
	\item Visualization of the inter-index association
	\item Visualization of drought index time series for selected locations
	\item Comparison of different variable orders for the calculation of multivariate drought indices
	\item Effect of trends on multivariate drought indices
\end{enumerate}

\subsection{Software and data}\label{suppB}

Moreover, we provide an \texttt{R} software package (\texttt{SIndices}, version 1.0) which is an implementation of the presented methodology. It comes along with a detailed manual. Further, we provide the \texttt{R}-code which was used to produce all results presented in the article and the supporting information. The Climatic Research Unit (CRU) time series (TS) data \citep[version 3.21, see][]{cru13} on which all examples and computations are based can be obtained from \url{http://dx.doi.org/10.5285/D0E1585D-3417-485F-87AE-4FCECF10A992}.

\section*{Acknowledgments}

The first author was supported by the Deutsche Forschungsgemeinschaft (DFG) through the TUM International Graduate School of Science and Engineering (IGSSE).
All computations were performed using the software environment \texttt{R} \citep[][]{R15}. To load the CRU data set we used the \texttt{raster} package \citep[][]{raster15}. To handle spatial and spatio-temporal data we used the packages \texttt{sp} \citep[][]{sp05} and \texttt{spacetime} \citep[][]{spacetime12}, respectively. To work with time series we used the package \texttt{xts} \citep[][]{xts14}. To calculate SPI and SPEI we used the \texttt{SPEI} package  \citep[][]{SPEI13}. For dependency modeling we used the \texttt{VineCopula} package \citep[][]{VineCopula15}. Empirical skewness estimates were calculated using the package \texttt{moments} \citep[][]{moments15}. For the Yeo and Johnson transformation we used the package \texttt{car} \citep[][]{car11}.

\bibliographystyle{chicago}	
\bibliography{SMVDI}

\end{document}

%% file: Sections/Introduction.tex
\section[Introduction]{Introduction}\label{sec:intro}

The challenging field of drought research has a long history. Scientists of different disciplines described and defined different drought concepts and tried to measure, quantify and predict drought events and their impacts. There exist several review papers trying to depict/portray the state of the art and different developments in drought modeling. One of the most recent and comprehensive ones is the review of drought concepts by \citet[][]{mishra10}. They state that ``drought is best characterized by multiple climatological and hydrological parameters''. Different drought types like \emph{meteorological drought} (lack of precipitation), \emph{hydrological drought} (declining water resources), \emph{agricultural drought} (lack of soil moisture), \emph{socio-economic drought} (excess demand for economic good(s) due to shortfall in water supply) or \emph{ground water drought} (decrease in groundwater recharge, levels and discharge) are driven by different variables/phenomena. Recently, there have been several attempts to develop \emph{multivariate drought indicators} \citep[see e.g.][]{kao10, hao13, hao14, farahmand15}, combining at least two different variables. Subsequently, we motivate and present a statistically sound approach for the calculation of standardized uni- and multivariate drought indices for arbitrary (sets of) drought relevant variables. The multivariate indices use so called vine copulas to flexibly model the variable dependencies.

Copulas are explained best by Sklar's Theorem \citep[][]{sklar59}. Let $F$ be a multivariate ($d$-dimensional) distribution function and $F_1,\ldots,F_d$ the corresponding marginals. Then there exists a copula $C$, such that $F(\bx)=C\left(F_1(x_1),\ldots,F_d(x_d)\right)$, where $\bx=(x_1,\ldots,x_d)\T$ is the realization of a (continuous) random vector $\bX\in\R^d$. A copula itself is a $d$-dimensional distribution function on the unit hypercube with uniformly distributed margins.
It captures all dependency information between the marginals of the corresponding multivariate distribution function. Vine copulas are $d$-dimensional copula constructions built on bivariate copulas only \citep[see][]{aas09, dissmann13}. They allow very flexible modeling of non-Gaussian, asymmetric dependency structures due to their modularity.

The most popular drought indices are the Palmer Drought Severity Index (\textbf{PDSI}) \citep[][]{palmer65} respectively its self-calibrating version (\textbf{SC-PDSI}) \citep[][]{wells04} and the Standardized Precipitation Index (\textbf{SPI}) \citep[][]{mckee93, edwards97}. Drought indices in general should quantify deviations from normal conditions, i.e. they should take \emph{seasonality} into account. Often \emph{negative/small values reflect dry conditions and positive/high values wet conditions}. They usually \emph{require long data records} to yield meaningful results.

The \emph{PDSI} is calculated based on precipitation and temperature and assumes a simplifying water balance model \citep[for details see][]{palmer65}. The major criticisms on the PDSI are its lack of applicability and comparability for different climatic regions. Some of its major shortcomings vanished with the \emph{SC-PDSI}, whose parameters are determined based on local climatic conditions rather than on some fixed locations in the US, i.e. it allows for \emph{spatial comparison}. One further criticism of the PDSI is its \emph{autoregressive structure}. Present conditions depend on past conditions, however the time interval which influences the present varies across space but cannot be accessed from the model.

In contrast to the PDSI, other drought indices like the \emph{SPI} \citep[][]{mckee93, edwards97} are of \emph{probabilistic nature}. This allows risk analysis, classification and frequency analysis of drought events. Two advantages of the purely precipitation based SPI over the PDSI are its \emph{standardization} (standard normal distribution of SPI values) and the concept of \emph{time scales}, which allows to set the time interval which has an influence on the present (drought) conditions. The SPI methodology \emph{can be applied to other variables} as well \citep[see e.g. the Standardized Runoff Index of][]{shukla08} and the standardization allows for \emph{comparison of such standardized indices and across space and time}. A criticism is that the SPI assumes a \emph{parametric distribution} to model the data. However, a good fit to the data (especially in the distribution tails) is never guaranteed and in fact is not possible for many locations (e.g. in the Sahara). Moreover, \emph{temporal dependencies} in the data or those introduced through the time scale cause the fitting to be biased.

As an enhancement of the SPI the Standardized Precipitation Evapotranspiration Index (\textbf{SPEI}) \citep[][]{vincente10} quantifies drought based on \emph{multivariate input}. Instead of precipitation a climatic water balance (precipitation minus potential evapotranspiration) is considered to quantify dry/wet conditions.
The SPEI allows for \emph{trends} in the time series data such that these are passed on to the index (to include effects of climate change).

\citet[][]{kao10} present a (to our knowledge) first multivariate copula-based drought index, the Joint Deficit Index (\textbf{JDI}). They apply it to precipitation and streamflow time-series, but application to other variables is possible. Marginals are modeled using the SPI approach. Empirical copulas are used to (non-parametrically) estimate the dependence structure of the marginals representing the different time scales of one to twelve months. Finally, the joint deficit index combines the drought information captured by different time scales using the Kendall distribution function to assess the joint probability. The results are transformed to a standard normal distribution. Note, that for meaningful estimation of empirical copulas long data records are required.

\citet[][]{farahmand15} introduce the Standardized Drought Analysis Toolbox (\textbf{SDAT}), with the aim to provide a generalized approach to derive non-parametric standardized drought indices. Based on precipitation and soil moisture time series, they present a multivariate approach to drought modeling. Enhancing the SPI idea to \emph{bivariate data} (based on non-parametric estimation), a bivariate empirical distribution is fitted to the input data and the joint cumulative probability is transformed with the inverse CDF of a standard normal distribution. Note however, that \emph{this approach doesn't yield a real standardization}. Usually negative values of the proposed index are more probable, since the joint cumulative probability is not uniformly distributed on $[0,1]$.

Summarizing the lessons learned from the sophisticated drought indices revised above, we state that (univariate) drought indices should \ldots
\begin{itemize}[leftmargin=4em]
	\item[\texttt{PROBAB}] be probabilistic (allow risk/frequency analysis and classification of drought events), i.e. no assumptions about the characteristics of the underlying system have to be made.
	\item[\texttt{ARBVAR}] be applicable to \emph{arbitrary drought relevant variables}.
	\item[\texttt{DRYWET}] be negative/positive to indicate \emph{dry/wet conditions}.
	\item[\texttt{SMALLS}] yield meaningful results for (monthly) data records for 10 years and more (i.e. \emph{minimum sample size} $=120$).
	\item[\texttt{TRENDS}] reflect \emph{trends} in the input data.
	\item[\texttt{SEASON}] model and eliminate \emph{seasonality}.
	\item[\texttt{TIMDEP}] model and eliminate \emph{temporal dependencies} before a probability distribution is fitted.
	\item[\texttt{NPDIST}] use \emph{non-parametric distribution estimates} for the (transformed) underlying variable (better fit, computationally efficient).
	\item[\texttt{STCOMP}] be \emph{standardized} to enable comparison over space/time and with other indices.
	\item[\texttt{TSCALE}] allow for computation/aggregation at different \emph{time scales} $l$.
	\item[\texttt{MULTEX}] be \emph{extendable to multivariate input} (different types of drought).
\end{itemize}
Table \ref{tab:modcomp} summarizes different drought indices and lists which characteristics they fulfill. Subsequently, we introduce a novel approach to drought modeling which addresses the above criteria step by step.

\begin{table}
\caption{\label{tab:modcomp}Comparison of different drought indices (SC-PDSI, SPI/SPEI, JDI, SDAT) and their properties: $+$ has this property, $-$ doesn't have this property, $?$ no definite answer possible or not applicable (e.g. because the corresponding model is not probabilistic).}
\centering
\fbox{%
\begin{tabular}{lcccc}
 & SC-PDSI & SPI/SPEI & JDI & SDAT \\
\hline
\texttt{PROBAB} & $-$ & $+$ & $+$ & $+$ \\
\texttt{ARBVAR} & $-$ & $?$ & $+$ & $+$ \\
\texttt{DRYWET} & $+$ & $+$ & $+$ & $+$ \\
\texttt{SMALLS} & $-$ & $-$ & $-$ & $-$ \\
\texttt{TRENDS} & $+$ & $+$ & $+$ & $+$ \\
\texttt{SEASON} & $+$ & $+$ & $+$ & $-$ \\
\texttt{TIMDEP} & $?$ & $-$ & $+$ & $-$ \\
\texttt{NPDIST} & $?$ & $-$ & $-$ & $+$ \\
\texttt{STCOMP} & $?$ & $+$ & $+$ & $-$ \\
\texttt{TSCALE} & $-$ & $+$ & $-$ & $+$ \\
\texttt{MULTEX} & $-$ & $?$ & $+$ & $+$
\end{tabular}}
\end{table}

%% file: Sections/Data.tex
\section[Data]{Data}\label{sec:data}
\begin{figure}[!htb]
	\centering
		\includegraphics[width=0.615\textwidth]{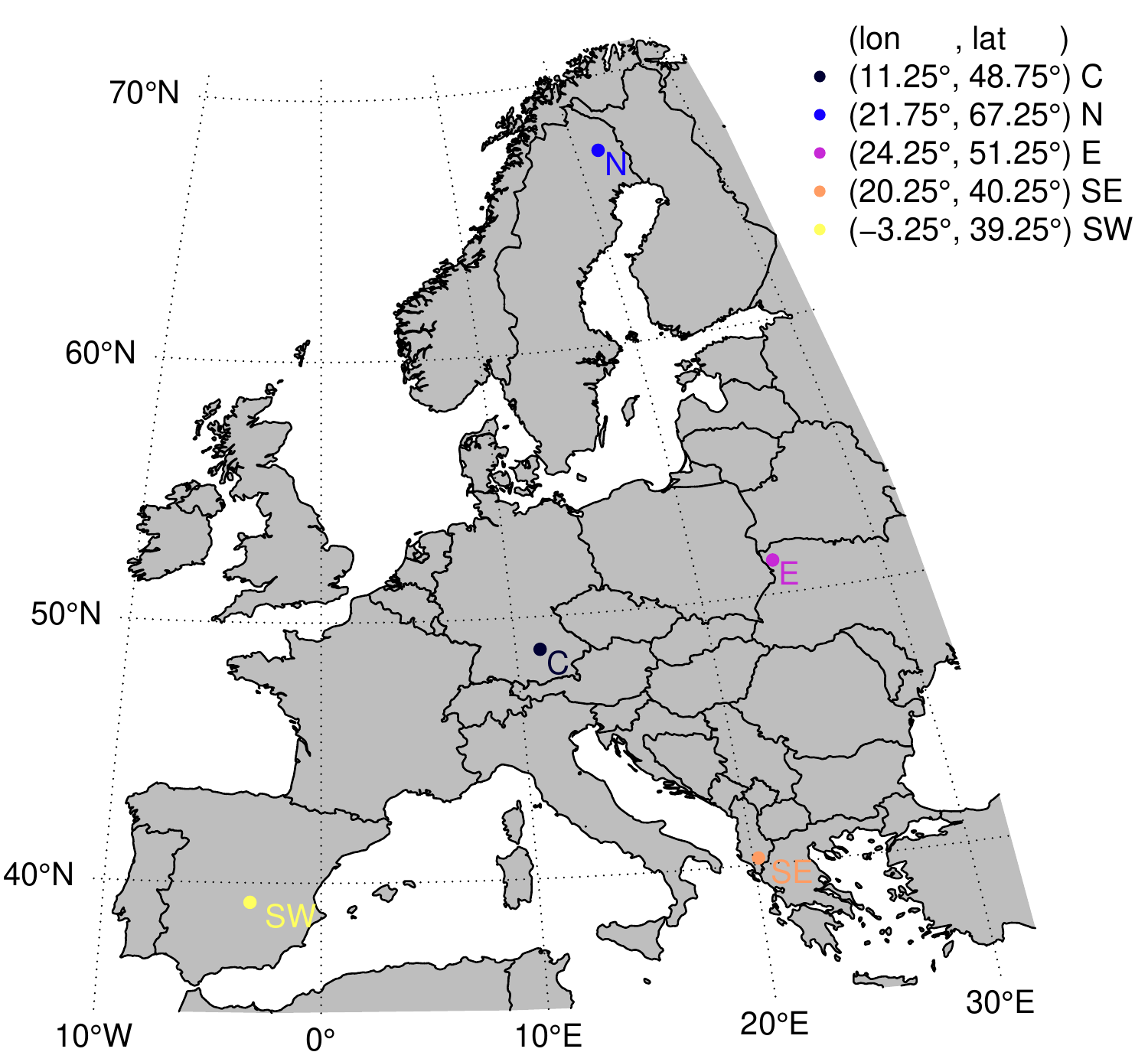}
			\caption{Study area and location of pixels used for illustration.}
		\label{fig:loc}
\end{figure}

For the purpose of application and illustration we utilize the publicly available Climatic Research Unit (CRU) time series (TS) data \citep[version 3.21, see][]{cru13}, which is monthly climatic data on a 0.5$^{\circ}\times$0.5$^{\circ}$ (longitude $\times$ latitude) grid. We restrict this (model-calculated) data set to the area $(11^{\circ}W, 32^{\circ}E)\times(35^{\circ}N, 71^{\circ}N)$ covering most of Europe (see gray shaded area in Figure \ref{fig:loc}). This results in data for $S=3380$ grid cells. For the calculation of drought indices we use the variables \emph{potential evapotranspiration} (\texttt{PET}), \emph{precipitation} (\texttt{PRE}) and \emph{vapor pressure deficit} (\texttt{VPD}) for the years 1961 to 2010 ($T=600$ months). \texttt{VPD} is calculated based on mean temperature (\texttt{TMP}) and vapor pressure (\texttt{VAP}) as $\text{\texttt{VPD}} = \text{\texttt{SVP}} - \text{\texttt{VAP}}$, where $\text{\texttt{SVP}} = 6.1078 \cdot 10^{\left[(7.5 \cdot \text{\texttt{TMP}})/ (\text{\texttt{TMP}} + 237.3)\right]}$ is the saturated vapor pressure \citep[see][]{murray67}. The five pixels C, N, E, SE and SW highlighted in Figure \ref{fig:loc} are used for subsequent illustrations. Their coordinates are provided in the figure. Time series plots corresponding to these five locations of the variables \texttt{PET}, \texttt{PRE}, \texttt{VPD} as well as SPI and SPEI are provided in the supporting information (see Section \ref{suppA}).

%% file: Sections/Univariate.tex
\section[Univariate standardized indices]{Univariate standardized indices}\label{sec:uni}

In a first step we seek to develop and illustrate a statistically sound (\texttt{PROBAB}) generalized modeling framework for (univariate) standardized indices. These indices should have the properties which were discussed in the introduction (Section \ref{sec:intro}).

\subsection{Variable transformation}\label{sec:trafo}
Let us now consider a time series $x_{t_k}$, $k=1,\ldots,T$, for an arbitrary drought relevant variable (\texttt{ARBVAR}). Small values should always indicate dry and big values wet conditions (\texttt{DRYWET}). To ensure that, we change the sign of the time series beforehand if it was the other way round. Consider for instance potential evapotranspiration (\texttt{PET}). A high value corresponds to potentially high evaporation and transpiration, i.e. to dry conditions. For low values the opposite is observed. Therefore we need to multiply the \texttt{PET} (and also the \texttt{VPD}) time series by $-1$.

\begin{figure}[!htb]
	\centering
		\includegraphics[width=1.00\textwidth]{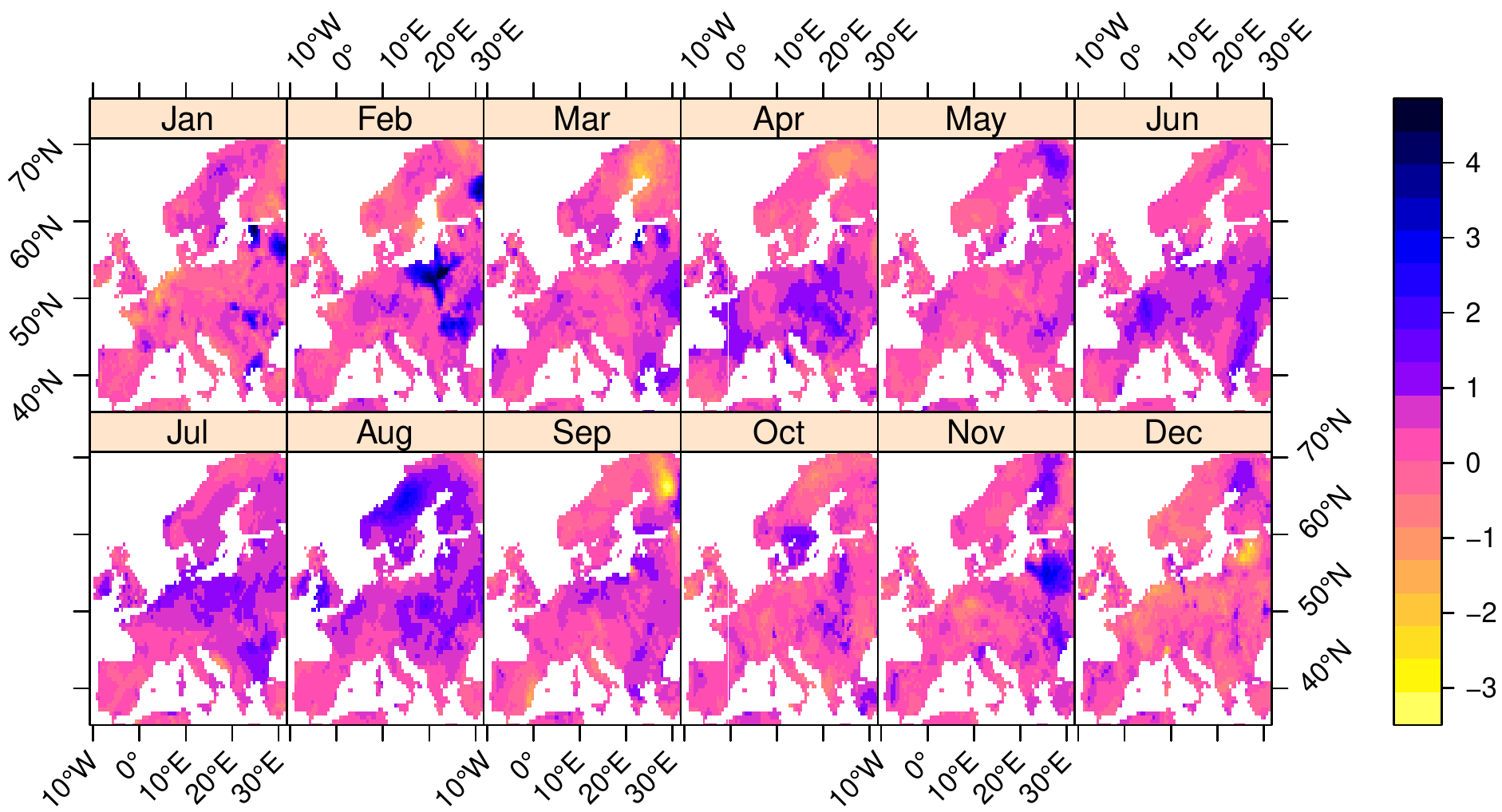}
	\caption{Empirical skewness estimates for the month-wise vapor pressure deficit (\texttt{VPD}) time series.}
	\label{fig:skewness}
\end{figure}

Subsequent steps include a month-wise standardization of the time series. Hence, it is preferable that the distribution of the time series for each month is not skewed. To achieve that, we consider monotone and continuous transformations. Figure \ref{fig:skewness} shows the spatial variation of skewness for the month-wise time series of vapor pressure deficit (\texttt{VPD}). We observe negative and positive skewness and a variation over the year, which supports a month-wise modeling approach.

Let each time point $t_k$, $k=1,\ldots,T$, be a $2$-tupel $(m_k,y_k)$, where $m_k \in \left\{1,\ldots,12\right\}$ ($1=$ January, \ldots, $12=$ December) represents the month and the integer $y_k \in \Z$ the year corresponding to $t_k$.
Then we consider the month-wise time series $\bx_m \coloneqq (x_{t_k})_{k\in\calK(m)} = \left\{x_{(m,y_k)}, k\in\calK(m)\right\}$, $m=1,\ldots,12$, where the index set for month $m$ is defined as $\calK(m) \coloneqq \left\{k:\, m_k = m\right\}$.

To eliminate/reduce skewness in the ($12$) month-wise time series $\bx_m$, $m=1,\ldots,12$, we apply power transformations.
An appropriate family of transformations, similar to the famous Box-Cox transformations, which is defined not only for positive values is the \citet[][]{yeo00} transformation $\psi:\R\times\R\to\R$, defined as
\beqo
\psi\left(\lambda,x\right) = \begin{dcases*}
\left((x+1)^\lambda-1\right)/\lambda & if $x\geq0, \lambda\neq0$ \\
\ln(x+1) & if $x\geq0, \lambda=0$ \\
-\left((-x+1)^{2-\lambda}-1\right)/(2-\lambda) & if $x<0, \lambda\neq2$ \\
-\ln(-x+1) & if $x<0, \lambda=2$.
\end{dcases*}
\eeqo

\begin{figure}[!htb]
	\centering
		\includegraphics[width=1.00\textwidth]{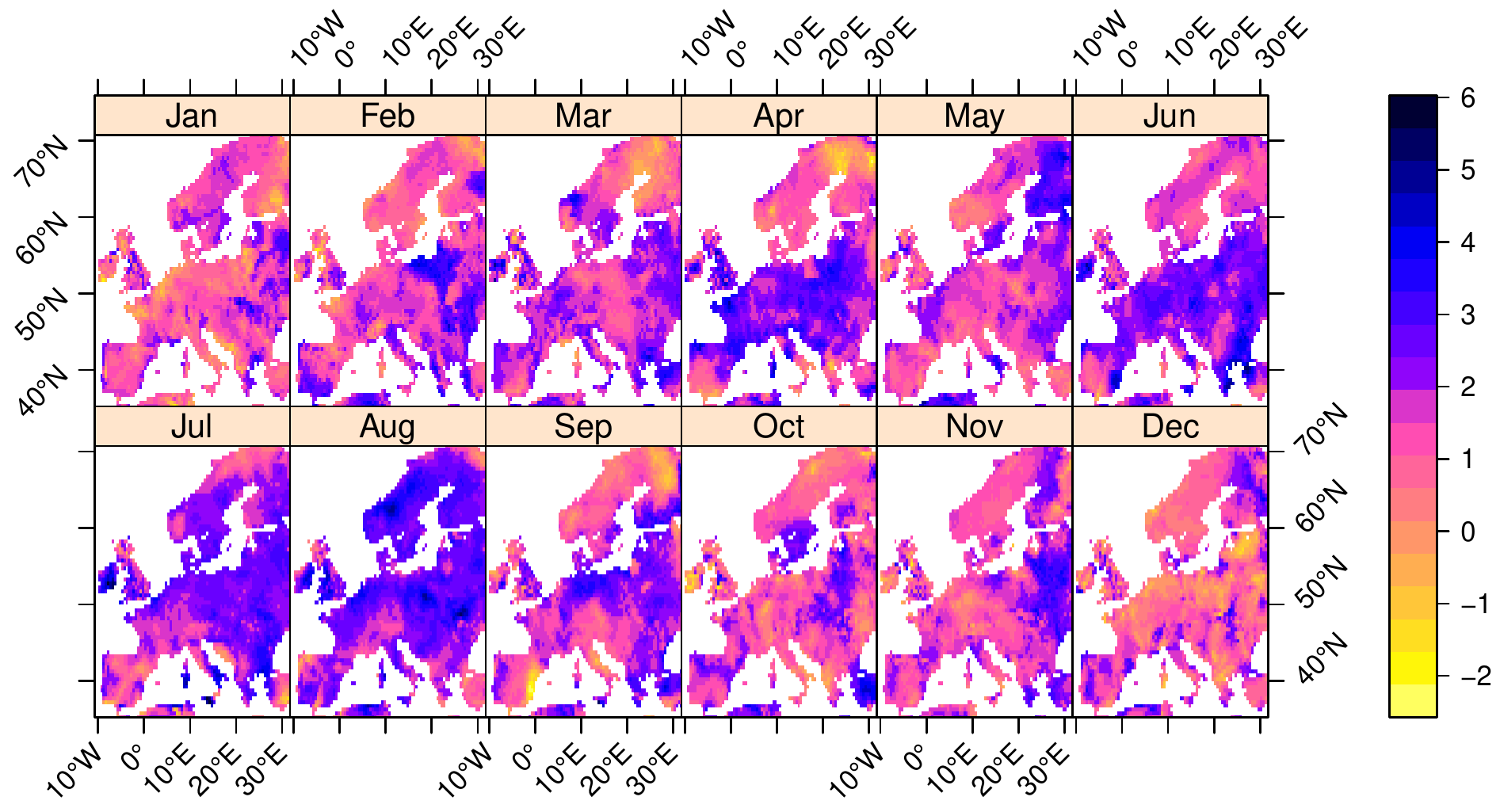}
	\caption{Yeo and Johnson transformation parameter $\lambda$ for the month-wise \texttt{VPD} time series.}
	\label{fig:lambda}
\end{figure}

Figure \ref{fig:lambda} maps the Yeo and Johnson transformation parameter $\lambda$ for the month-wise \texttt{VPD} time series. The observed spatial paterns resemble those observed for skewness in Figure \ref{fig:skewness}.

\subsection{Elimination of seasonality}\label{sec:seas}
\begin{figure}[!htb]
	\centering
		\includegraphics[width=1.00\textwidth]{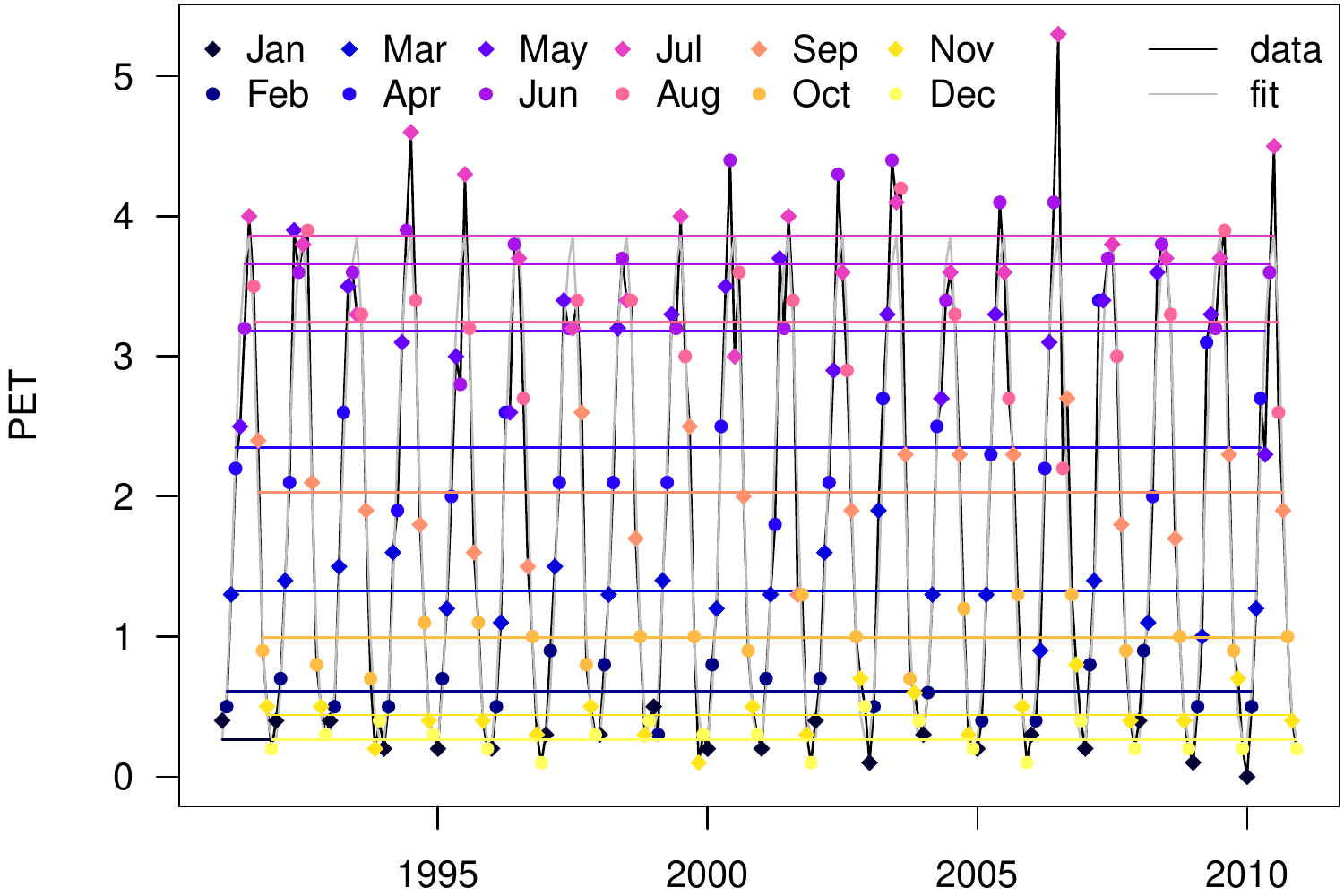}
	\caption{Month-wise modeling of \texttt{PET}: The original time series (black, $1991$-$2010$, for pixel C as given in Figure \ref{fig:loc}) is superimposed by the corresponding month-wise mean fit (grey, cp. Equation \eqref{eq:mean}). The month-wise time series are illustrated by points colored differently for each month. The modeled month-wise means are visualized by lines in the corresponding color.}
	\label{fig:trends}
\end{figure}

Often (climatic) variables are subject to seasonal fluctuations (see e.g. \texttt{PET}, Figure \ref{fig:trends}). Moreover, they can be subject to trends (e.g. due to climate change). \texttt{TRENDS} are not removed since a drought index should be able to detect changes in drought frequency and intensity due to climate change. Since drought is considered as a (negative) deviation from `normal' conditions (anomaly), we remove seasonality (\texttt{SEASON}). This is accounted for by month-wise modeling of the time series $x_{t_k}$, $k=1,\ldots,T$. However, to ensure that the sample size (\texttt{SMALLS}) for fitting a distribution is not too small, our deseasonalization procedure allows to recompose the resulting anomalies to a single time series.

To eliminate seasonality, we model the month-wise mean $\mu_m$ separately for each of the $12$ time series $\bx_m$, $m=1,\ldots,12$. We estimate it as
\beqq\label{eq:mean}
	\wh\mu_m \coloneqq \frac{1}{|\calK(m)|}\sum_{k\in\calK(m)}x_{(m,y_k)}, \quad m=1,\ldots,12.
\eeqq
Figure \ref{fig:trends} illustrates the month-wise modeling \eqref{eq:mean} of potential evapotranspiration (\texttt{PET}). 
Least-squares estimation ensures that $\sum_{k\in\calK(m)}\left(x_{(m,y_k)}-\wh\mu_m\right)=0$ for all $m=1,\ldots,12$. Thus also the anomalies $a_{t_k} \coloneqq x_{t_k} - \wh\mu_{m_k}$, $k=1,\ldots,T$, are centered around $0$ (i.e. $\sum_{k=1}^{T}a_{t_k}=0$). Hence, seasonal deviations from the annual mean could be eliminated.

Also the variance of the time series may be subject to seasonality, i.e. in some months the time series may deviate more from its mean compared to other months. The color-coding in Figure \ref{fig:trends} reveals inhomogeneity of the variance. To quantify this seasonal heterogeneity of the time series $a_{t_k}$, $k=1,\ldots,T$, we estimate month-wise standard deviations as
\beqo
	\wh\si_m \coloneqq \sqrt{\frac{1}{|\calK(m)|-1}\sum_{k\in\calK(m)}a_{(m,y_k)}^2}, \quad m=1,\ldots,12,
\eeqo
where $|\cdot|$ is the cardinality. To obtain a homogenized time series we compute the standardized anomalies (residuals) $r_{t_k} \coloneqq a_{t_k}/\wh\si_{m_k}$, $k=1,\ldots,T$.

\subsection{Elimination of temporal dependencies}\label{sec:time}
Apart from seasonality, time series often feature temporal dependence (\texttt{TIMDEP}). Such serial dependencies can be captured by \emph{autoregressive moving-average models} (see e.g. \cite{box08}). For a (deseasonalized, homogeneous, zero-mean) time series $r_{t_k}$, $k=1,\ldots,T$, the autoregressive moving-average model ARMA($p$, $q$) with AR-order $p\in\N_0$ and MA-order $q\in\N_0$ is defined as
\beqo
	r_{t_k} = \sum_{j=1}^{p}\phi_{j}r_{t_{k-j}} + \sum_{j=1}^{q}\te_{j}\veps_{t_{k-j}} + \veps_{t_k},
\eeqo
where the error terms $\veps_{t_k}$ are i.i.d. $N(0,\si^2)$ distributed. Note, that for $p$ or $q$ equal to $0$ the corresponding summands are neglected. For adequate choice of the orders $p$ and $q$ and estimates $\wh\phi_{j}$, $j=1,\ldots,p$, and $\wh\te_{j}$, $j=1,\ldots,q$, of the corresponding parameters the model residuals
$\eps_{t_k} \coloneqq r_{t_k} - \sum_{j=1}^{p}\wh\phi_{j}r_{t_{k-j}} - \sum_{j=1}^{q}\wh\te_{j}\eps_{t_{k-j}}$, $k=1,\ldots,T$, 
are approximately temporally independent. For the variables at hand (\texttt{PET}, \texttt{PRE}, \texttt{VPD}) $p=1$ and $q=0$ are an adequate choice.

\subsection{Transformation to standard normal distribution}\label{sec:zscale}
As the assumption of established standardized drought indices like SPI and SPEI of a parametric distribution model for the data performs bad, it seems appropriate to use the (non-parametric) empirical distribution (\texttt{NPDIST}) function $\wh F_T(x) \coloneqq \frac{1}{T}\sum_{k=1}^{T}\ind{\eps_{t_k} \leq x}$ of the data respectively the residuals $\eps_{t_k}$, $k=1,\ldots,T$, resulting from the previous modeling step. Here $\ind{\calA}$ is the indicator function, which equals $1$ if the event $\calA$ is true and $0$ otherwise. Note that for fitting a distribution (no matter if parametric or not) to a sample $\eps_{t_k}$, $k=1,\ldots,T$, it is a critical assumption that the sample originates from the same distribution and is i.i.d. We ensured the i.i.d. assumption in the previous step by eliminating the temporal dependencies.

We use the estimated distribution $\wh F_T$ to transform our residuals $\eps_{t_k}$, $k=1,\ldots,T$, to the u-scale, i.e. to be (approximately) uniformly distributed on the interval $[0,1]$. This transformation is called probability integral transform (PIT). We calculate $u_{t_k} \coloneqq T/(T+1) \wh F_T\left(\eps_{t_k}\right) = \rank\left(\eps_{t_k}\right)/(T+1)$, $k=1,\ldots,T$. We multiply by $T/(T+1)$ to avoid any $u_{t_k}=1$. Further, we transform to the z-scale, calculating $z_{t_k} \coloneqq \Phi^{-1}\left(u_{t_k}\right)$, $k=1,\ldots,T$, using the inverse PIT based on the CDF $\Phi$ of a standard normal distribution. It holds that $z_{t_k}$, $k=1,\ldots,T$, is (approximately) independent and identically standard normal distributed (\texttt{STCOMP}).

\subsection{Standardized indices on different time scales}\label{sec:tscale}
\citet[][]{mckee93} introduced the concept of time scales (\texttt{TSCALE}) to make their drought index (the SPI) applicable to different types of drought. We adopt this concept, however we perform the temporal aggregation in the end of the above described modeling process, in order not to violate the independency assumption for fitting a probability distribution to the residuals. This has also the advantage of being computationally more efficient. We need to perform the different modeling steps of Sections \ref{sec:trafo}-\ref{sec:zscale} only once, after that we are able to calculate the index on arbitrary time scales. 

The (approximately) temporally independent standard normal distributed time series $z_{t_k}$, $k=1,\ldots,T$, from above is already a standardized index with time scale $l=1$. The normal distribution has the advantage that a sum of independent normal distributed random variables is again normally distributed. We use this property to calculate standardized indices for time scales $l \geq 1$. The sum $\sum_{j=1}^{l}z_{t_{k+1-j}}$ of standard normal variables is normally distributed with mean $0$ and variance $l$. Hence, we obtain a \emph{standardized index} with time scale $l$ as $\SI_l(t_k) \coloneqq \frac{1}{\sqrt{l}}\sum_{j=1}^{l}z_{t_{k+1-j}}$, $k=1,\ldots,T$.

\begin{table}
\caption{\label{tab:cat}Dryness and wetness categories.}
\centering
\fbox{%
\begin{tabular}{r|rcc}
&	&	cumulative &	\\
&	category	&	probability	&	quantile	\\
\hline
$W4$	&	exceptionally wet	&	$0.98$-$1.00$	&	$+2.05<\SI<+\infty$	\\
$W3$	&	extremely wet	&	$0.95$-$0.98$	&	$+1.64<\SI\leq+2.05$	\\
$W2$	&	severely wet	&	$0.90$-$0.95$	&	$+1.28<\SI\leq+1.64$	\\
$W1$	&	moderately wet	&	$0.80$-$0.90$	&	$+0.84<\SI\leq+1.28$	\\
$W0$	&	abnormally wet	&	$0.70$-$0.80$	&	$+0.52<\SI\leq+0.84$	\\
$D0$	&	abnormally dry	&	$0.20$-$0.30$	&	$-0.84<\SI\leq-0.52$	\\
$D1$	&	moderately dry	&	$0.10$-$0.20$	&	$-1.28<\SI\leq-0.84$	\\
$D2$	&	severely dry	&	$0.05$-$0.10$	&	$-1.64<\SI\leq-1.28$	\\
$D3$	&	extremely dry	&	$0.02$-$0.05$	&	$-2.05<\SI\leq-1.64$	\\
$D4$	&	exceptionally dry	&	$0.00$-$0.02$	&	$-\infty<\SI\leq-2.05$
\end{tabular}}
\end{table}

\begin{figure}[!htb]
	\centering
		\includegraphics[width=1.00\textwidth]{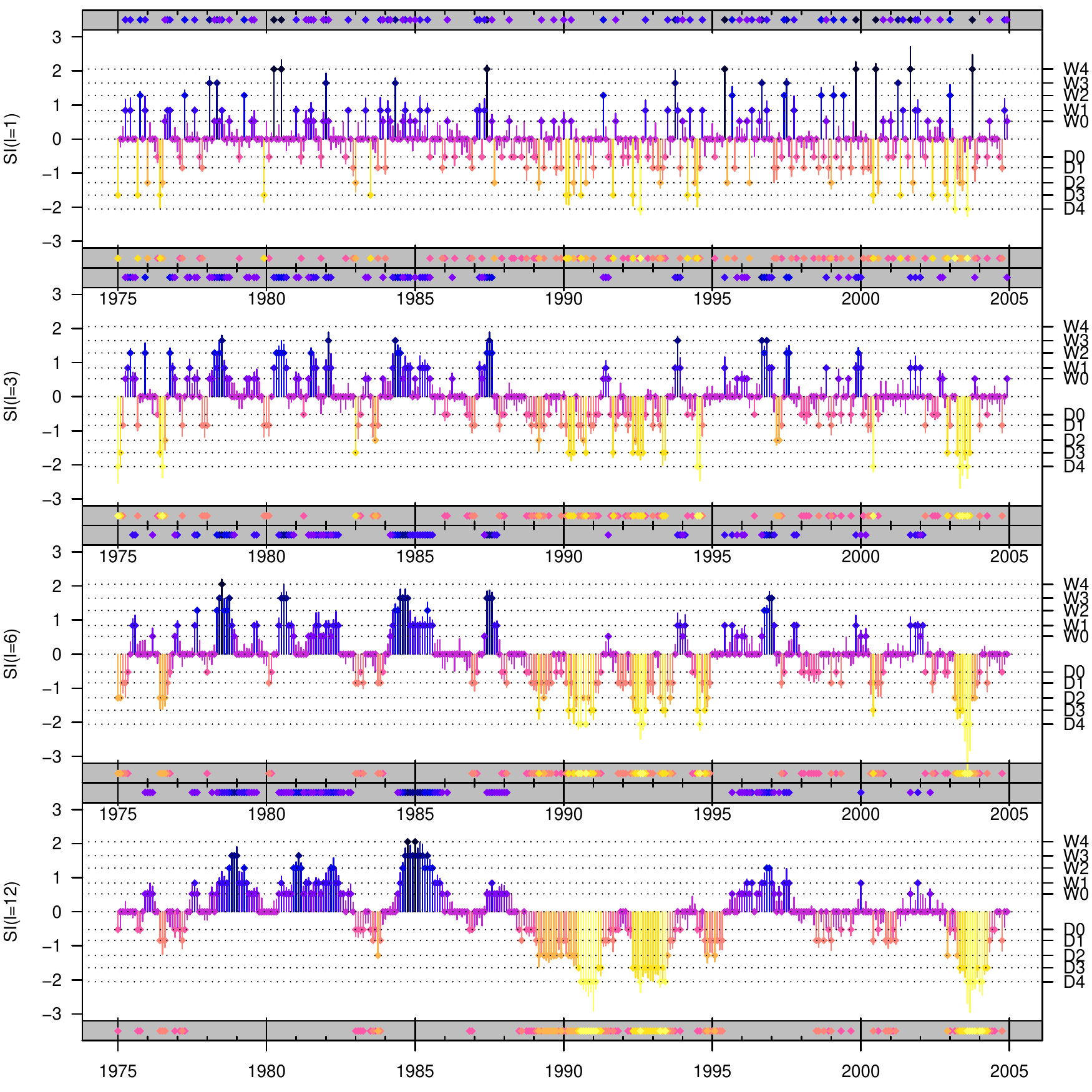}
	\caption{Time series ($1975$ - $2004$, for pixel C Figure \ref{fig:loc}) of standardized drought index (SI) based on \texttt{PRE}, for time scales $l=1, 3, 6$ and $12$, repectively. The color-coding reflects the severity of wetness/dryness according to the different categories specified in Table \ref{tab:cat}. For better identification of dry/wet periods points at the top/bottom of the panels (colored accordingly) indicate points in time of wet/dry conditions.}
	\label{fig:scales}
\end{figure}

To classify the values of standardized indices we use the dryness/wetness categories as defined in Table \ref{tab:cat} based on quantiles \citep[cp.][]{svoboda02}. A comparison of precipitation (\texttt{PRE}) based drought indices for different time scales is provided by Figure \ref{fig:scales}. For the selected location we identify persistent dry periods during the years $1976$, $1989-1991$, $1992-1993$ and $2003-2004$. Whereas the index with time scale $1$ identifies single (agricultural) drought months, higher time scales (e.g. $6$, $12$) allow to identify persistent periods of dryness (hydrological drought).

%% file: Sections/Multivariate.tex
\section[Multivariate standardized indices]{Multivariate standardized indices}\label{sec:mult}

Subsequently, we provide an extension of the methodology introduced in Section \ref{sec:uni} to multivariate standardized drought indices (\texttt{MULTEX}). This extension is based on vine copulas \citep[see][]{aas09} used for dependency modeling of the involved variables. The dependence parameters will be estimated using a semi-parametric estimation procedure \citep[see][]{genest95}. Other copula based drought indices were introduced by \citet[][]{farahmand15} and \citet[][]{kao10}.

\subsection{Marginal models}\label{sec:margins}
As copulas allow separate modeling of margins and dependence structure, we first model the margins according to Sections \ref{sec:trafo}-\ref{sec:zscale} as in the univariate case. We transform the input data (see Section \ref{sec:trafo}), then we eliminate seasonality (see Section \ref{sec:seas}) and temporal dependencies (see Section \ref{sec:time}) and estimate the distribution of the remaining residuals non-parametrically (see Section \ref{sec:zscale}). This enables transformation to the u-scale (copula data) and after that copula based dependency modeling.

\subsection{Vine copula based dependency modeling}\label{sec:mdep}
Let now $\bu \coloneqq (\bu_1,\ldots,\bu_d)$ be the copula data obtained from the marginal models corresponding to $d$ different drought relevant variables, where $\bu_j = (u_{j,t_k})_{k=1,\ldots,T}$, $j=1,\ldots,d$, and $u_{j,t_k}$ is the copula data corresponding to variable $j$ at time $t_k$.
In a second (parametric) step we select and estimate a \emph{vine copula} $C$ for this data. We illustrate this procedure based on a $d=3$ dimensional example. For a more general explanation of vine copulas see \citet[][]{aas09} and \citet[][]{dissmann13}.

\begin{figure}[!htb]
	\centering
		\includegraphics[width=1.00\textwidth]{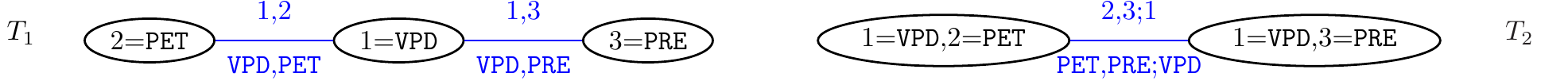}
	\caption{Selected vine tree structure.}
	\label{fig:cvine}
\end{figure}

Let now $d=3$ and $1=\text{\texttt{VPD}}$, $2=\text{\texttt{PET}}$ and $3=\text{\texttt{PRE}}$. Generally, the structure of vine copulas is organized using a nested set of trees (graphs) fulfilling certain conditions. The edges of these trees correspond to bivariate copulas which are the building blocks of the vine copula. Selecting the tree structure as given in Figure \ref{fig:cvine}, we explicitly model the bivariate dependence structures (copulas) $C_{1,2}$, $C_{1,3}$ (tree $T_1$) and $C_{2,3;1}$ (tree $T_2$) for the variable pairs (\texttt{VPD},\texttt{PET}), (\texttt{VPD},\texttt{PRE}) and (\texttt{PET},\texttt{PRE}) given \texttt{VPD}, respectively. Here $C_{2,3;1}$ denotes the pair-copula associated with the conditional distribution of the variable pair $(2,3)$ given variable $1$. Further, we select pair-copula families for the pairs above and denote their parameters as $\btheta \coloneqq (\theta_{1,2},\theta_{1,3},\theta_{2,3;1})$. Then the vine copula density $c$ is given as
\begin{align*}
	c(u_1,u_2,u_3; \btheta) = & \, c_{1,2}(u_1,u_2; \theta_{1,2}) \cdot c_{1,3}(u_1,u_3; \theta_{1,3}) \\
	& \cdot c_{2,3;1}(h_{2|1}(u_2, u_1; \theta_{1,2}),h_{3|1}(u_3, u_1; \theta_{1,2}); \theta_{2,3;1}),
\end{align*}
where $c_{1,2}$, $c_{1,3}$ and $c_{2,3;1}$ are the pair-copula densities corresponding to the copulas $C_{1,2}$, $C_{1,3}$ and $C_{2,3;1}$. The involved h-functions are defined as $h_{b|a}(u_b, u_a; \theta) \coloneqq C_{b|a}(u_b|u_a; \theta)$, where $C_{b|a}$ denotes the conditional distribution function of $U_b$ given $U_a$. The tree structure can be saved in a triangular, so called R-vine matrix. For the given three dimensional example a valid R-vine matrix is given as
\beqo
\left(\begin{matrix}
3 & 0 & 0 \\
2 & 2 & 0 \\
1 & 1 & 1 \\
\end{matrix}\right)
\quad\text{respectively}\quad
\left(\begin{matrix}
\text{\texttt{PRE}} & 0 & 0 \\
\text{\texttt{PET}} & \text{\texttt{PET}} & 0 \\
\text{\texttt{VPD}} & \text{\texttt{VPD}} & \text{\texttt{VPD}} \\
\end{matrix}\right).
\eeqo
Whereas the second column encodes the pair (\texttt{VPD},\texttt{PET}), the first column contains the pairs (\texttt{VPD},\texttt{PRE}) and (\texttt{PET},\texttt{PRE};\texttt{VPD}). Other orders of these variables are possible. For a comparison of different orders see the supporting information (Section \ref{suppA}).

For the pair-copula family selection we can choose among a variety of bivariate copula families, amongst others among the Gaussian (N), Student-$t$ (t), Clayton (C), Gumbel (G) Frank (F) and Joe (J) family, which all feature different dependence structures and properties. Also rotated versions of the Clayton, Gumbel and Joe copula are considered to capture negative asymmetric dependencies. The pair-copulas are selected separately (according to the BIC) starting in tree $T_1$. Their parameters are estimated at the same time using maximum likelihood estimation. Before that a bivariate independence test \citep[][]{genest07} can be performed, to see if an independence copula should be selected. For more details on different (rotated) copula families and their selection we refer to \citet[][]{cdvine13}.

\begin{figure}[!htb]
   \centering
    \begin{minipage}{.33\textwidth}
        \centering
        \includegraphics[width=\linewidth]{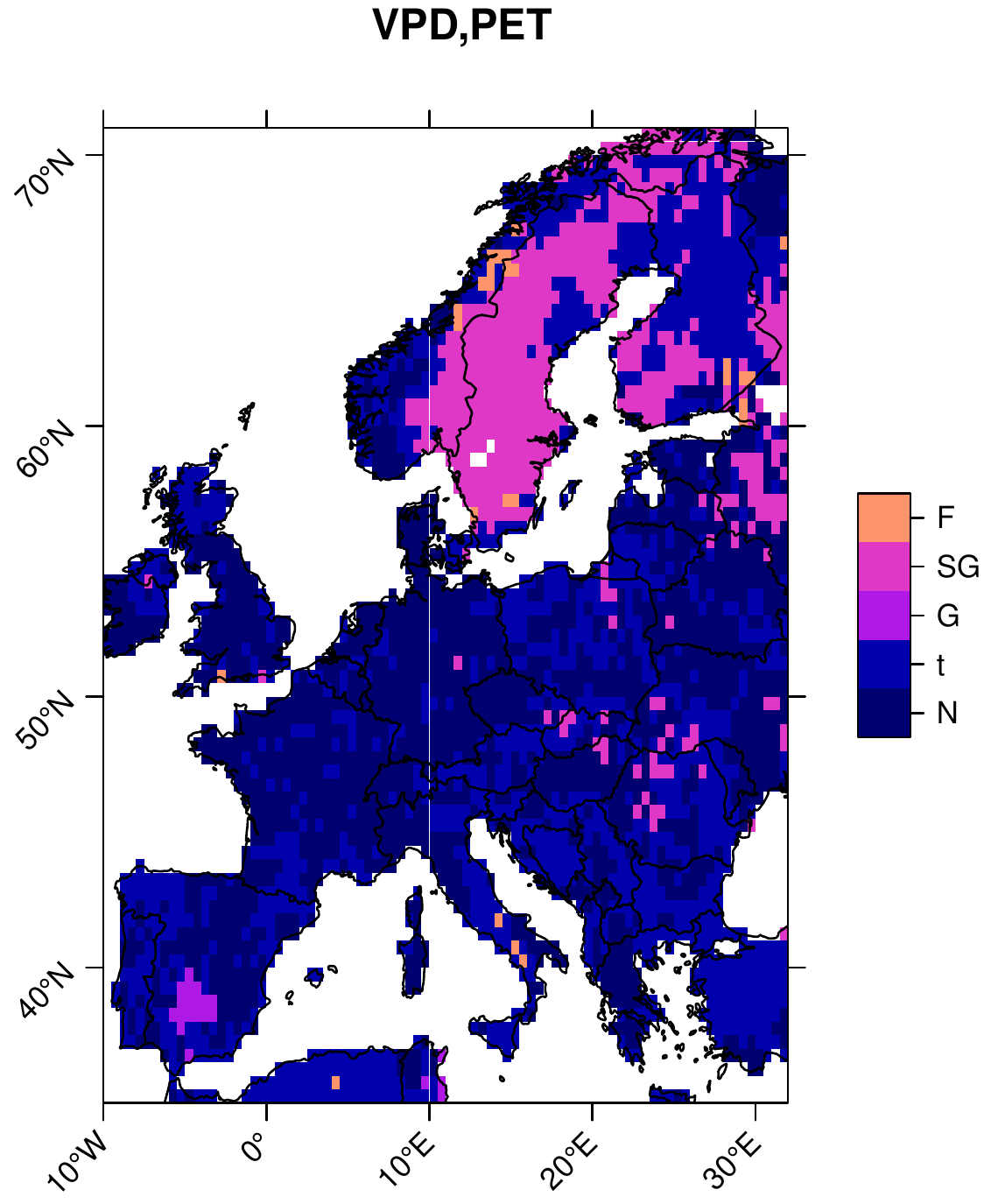}
    \end{minipage}%
    \begin{minipage}{0.33\textwidth}
        \centering
        \includegraphics[width=\linewidth]{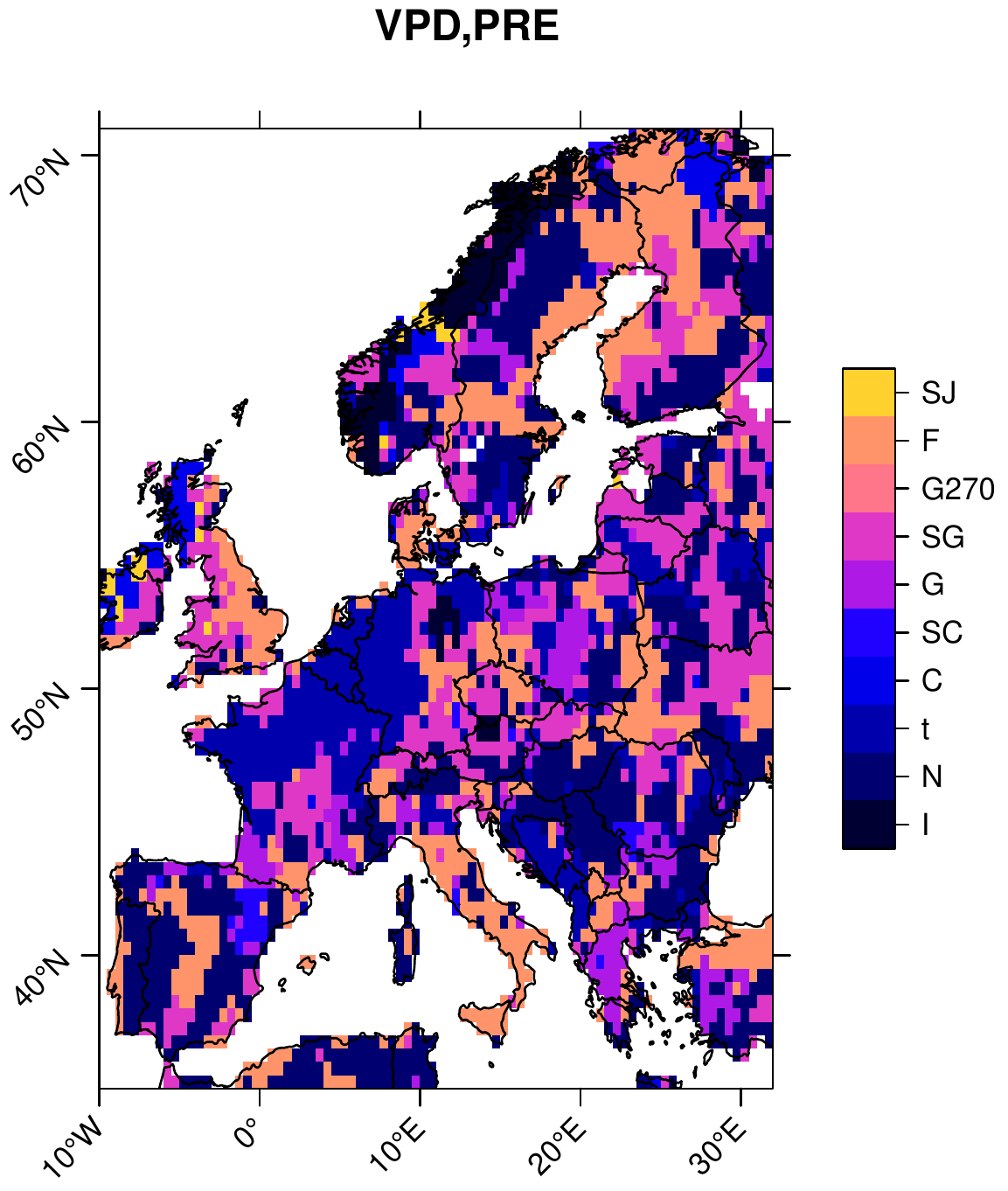}
    \end{minipage}
		\begin{minipage}{0.33\textwidth}
        \centering
        \includegraphics[width=\linewidth]{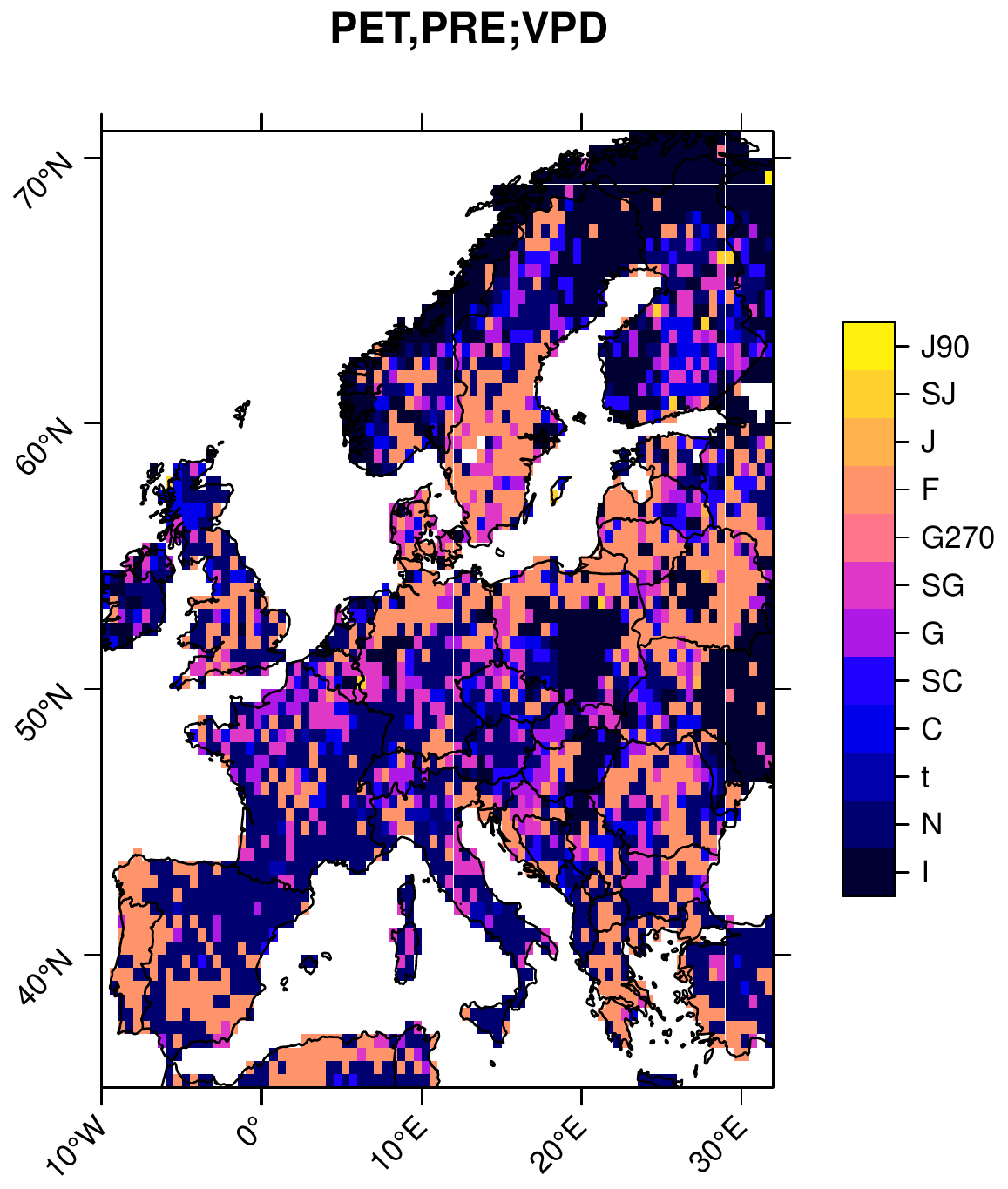}
    \end{minipage}
	\caption{Spatial variation in the pair-copula families selected for the pairs specified in Figure \ref{fig:cvine}.}
	\label{fig:vinestr}
\end{figure}

Figure \ref{fig:vinestr} visualizes for all spatial pixels under consideration which dependence structures were selected for the vine tree structure specified above (Figure \ref{fig:cvine}). For the pair (\texttt{VPD},\texttt{PET}) the elliptical and symmetric Gaussian (N) and Student-$t$ (t) copula were selected most over Europe. Where the Student-$t$ copula was selected extreme high or low \texttt{VPD} and \texttt{PET} anomalies occur jointly, since the Student-$t$ copula allows for dependence in the upper and lower distribution tails, so called tail-dependence. For a large area on the Iberian Peninsula (G) extreme wet conditions for both variable pairs seem to occur simultaneously, since the Gumbel copula allows for upper tail dependence. For most of Scandinavia (SG) we observe the opposite, high correlation of extreme dry conditions, since the survival/$180^{\circ}$ rotated Gumbel copula allows for lower tail dependence. For the other two (conditioned) pairs similar interpretations can be made. We observe that for most pixels non-Gaussian dependence structures were selected.

\subsection{Computation of multivariate indices}\label{sec:mind}
Based on the previously selected vine copula $C$ for the data $\bu = (\bu_1,\ldots,\bu_d)$, we transform $\bu$ to i.i.d. uniform data on $[0,1]$, using the so called \emph{\citet[][]{rosenblatt52} transformation}, a multivariate probability integral transform. The Rosenblatt transform $\bv \coloneqq (\bv_1,\ldots,\bv_d)$ of $\bu$ is defined as
\begin{align*}
	v_{1,t_k} & \coloneqq u_{1,t_k}, \\
	v_{2,t_k} & \coloneqq C_{2|1}(u_{2,t_k}|u_{1,t_k}), \\
	 & \ldots \\
	v_{d,t_k} & \coloneqq C_{d|1,\ldots,d-1}(u_{d,t_k}|u_{1,t_k},\ldots,u_{d-1,t_k}),\quad k=1,\ldots,T
\end{align*}
where $C_{j|1,\ldots,j-1}$, is the conditional cumulative distribution function for variable $j$ given the variables $1,\ldots,j-1$, for all $j = 2,\ldots,d$. For vine copulas the order of the variables is determined by the vine tree structure respectively the R-vine matrix. For details on the computation of the Rosenblatt transform for vine copulas see \citet[][]{schepsmeier15}.

Generally speaking, application of the Rosenblatt transform to our $d$ dependent variables yields independent information about dry/wet conditions captured in these variables. $\bv_1$ incorporates the same information as an univariate drought index calculated according to Section \ref{sec:uni} based on variable $1$. $\bv_j$, $j = 2,\ldots,d$, provide information on dry/wet conditions identified by variable $j$, conditioned on the dryness/wetness information provided by the previously considered variables $1,\ldots,j-1$.

For our three dimensional example from above we compute $v_{\text{\texttt{VPD}},t}=u_{\text{\texttt{VPD}},t}$, which represents the dry-/wetness information captured in the variable \text{\texttt{VPD}} for time point $t$. $v_{\text{\texttt{PET}},t} = C_{\text{\texttt{PET}}|\text{\texttt{VPD}}}(u_{\text{\texttt{PET}},t}|u_{\text{\texttt{VPD}},t})$ provides additional information based on \texttt{PET} knowing about \texttt{VPD} in that particular time point $t$. Its calculation involves the pair-copula $C_{\text{\texttt{VPD}},\text{\texttt{PET}}}$. The calculation of $v_{\text{\texttt{PRE}},t} = C_{\text{\texttt{PRE}}|\text{\texttt{VPD}},\text{\texttt{PET}}}(u_{\text{\texttt{PRE}},t}|u_{\text{\texttt{VPD}},t},u_{\text{\texttt{PET}},t})$ is a bit more involved. We calculate $v_{\text{\texttt{PRE}},t} = C_{\text{\texttt{PRE}}|\text{\texttt{PET}};\text{\texttt{VPD}}}(C_{\text{\texttt{PRE}}|\text{\texttt{VPD}}}(u_{\text{\texttt{PRE}},t}|u_{\text{\texttt{VPD}},t})|C_{\text{\texttt{PET}}|\text{\texttt{VPD}}}(u_{\text{\texttt{PET}},t}|u_{\text{\texttt{VPD}},t}))$, based on the pair-copulas $C_{\text{\texttt{PRE}},\text{\texttt{PET}};\text{\texttt{VPD}}}$, $C_{\text{\texttt{VPD}},\text{\texttt{PRE}}}$ and $C_{\text{\texttt{VPD}},\text{\texttt{PET}}}$.

Subsequently we consider two different approaches to join this multivariate drought information into one index. For comparison, we provide a third approach which assumes multivariate normality.
\paragraph{Method $\calA$ (aggregation)}
This approach allows for a weighting with weights $\bw = (w_1,\ldots,w_d)$, $w_j>0$, for the different variables $j=1,\ldots,d$. We calculate the \emph{standardized multivariate index (SMI)} with time scale $l$ as
\beqo
	\ASMI_l(\bw;1,\ldots,d)(t_k) \coloneqq \frac{1}{\sqrt{l \bw\T\bw}}\sum_{i=1}^{l}\sum_{j=1}^{d}w_j \Phi^{-1}\left(v_{j,t_{k+1-i}}\right).
\eeqo

\paragraph{Method $\calM$ (multiplication)}
For the second approach we exploit that the multivariate dependence structure of $\bv = (\bv_1,\ldots,\bv_d)$ is represented by the independence copula $C_\Pi(v_1,\ldots,v_d)=\prod_{j=1}^{d}v_j$. Hence, we calculate $\wt v_{t_k} \coloneqq \prod_{j=1}^{d}v_{j,t_k}$, $k=1,\ldots,T$. To obtain a standardized (multivariate) index we proceed as in the univariate case (see Sections \ref{sec:zscale} and \ref{sec:tscale}). We calculate the rank transformation $\wt u_{t_k} \coloneqq \rank\left(\wt v_{t_k}\right)/(T+1)$, $k=1,\ldots,T$, transform to the z-scale and calculate the SMI with time scale $l$ as
\beqo
	\MSMI_l(\beins;1,\ldots,d)(t_k) \coloneqq \frac{1}{\sqrt{l}}\sum_{i=1}^{l}\Phi^{-1}\left(\wt u_{t_{k+1-i}}\right),
\eeqo
where no weighting is allowed, i.e. $\bw=\beins\coloneqq(1,\ldots,1)$.

\paragraph{Method $\calN$ (normal)}
Let $\bz$ be the marginal transformation of $\bu$ to the z-scale and consider a vector of weights $\bw = (w_1,\ldots,w_d)$, $w_j>0$. Assuming $\bz$ to be a sample from a zero mean multivariate normal distribution, we can conclude that the linear transformation $\bw\T \bz$ is a sample from a zero mean univariate normal distribution. We estimate the sample variance of $\bw\T \bz$ by $S \coloneqq \frac{1}{T-1}\sum_{k=1}^{T}\left(\sum_{j=1}^{d}w_j \Phi^{-1}(u_{j,t_k})\right)^2$ and calculate a (weighted) SMI with time scale $l$ as
\beqo
	\NSMI_l(\bw;1,\ldots,d)(t_k) \coloneqq \frac{1}{\sqrt{l \cdot S}}\sum_{i=1}^{l}\sum_{j=1}^{d}w_j \Phi^{-1}\left(u_{j,t_{k+1-i}}\right).
\eeqo

%% file: Sections/Verification.tex
\section[Application]{Application}\label{sec:verifi}

\begin{figure}[!htb]
	\centering
		\includegraphics[width=1.00\textwidth]{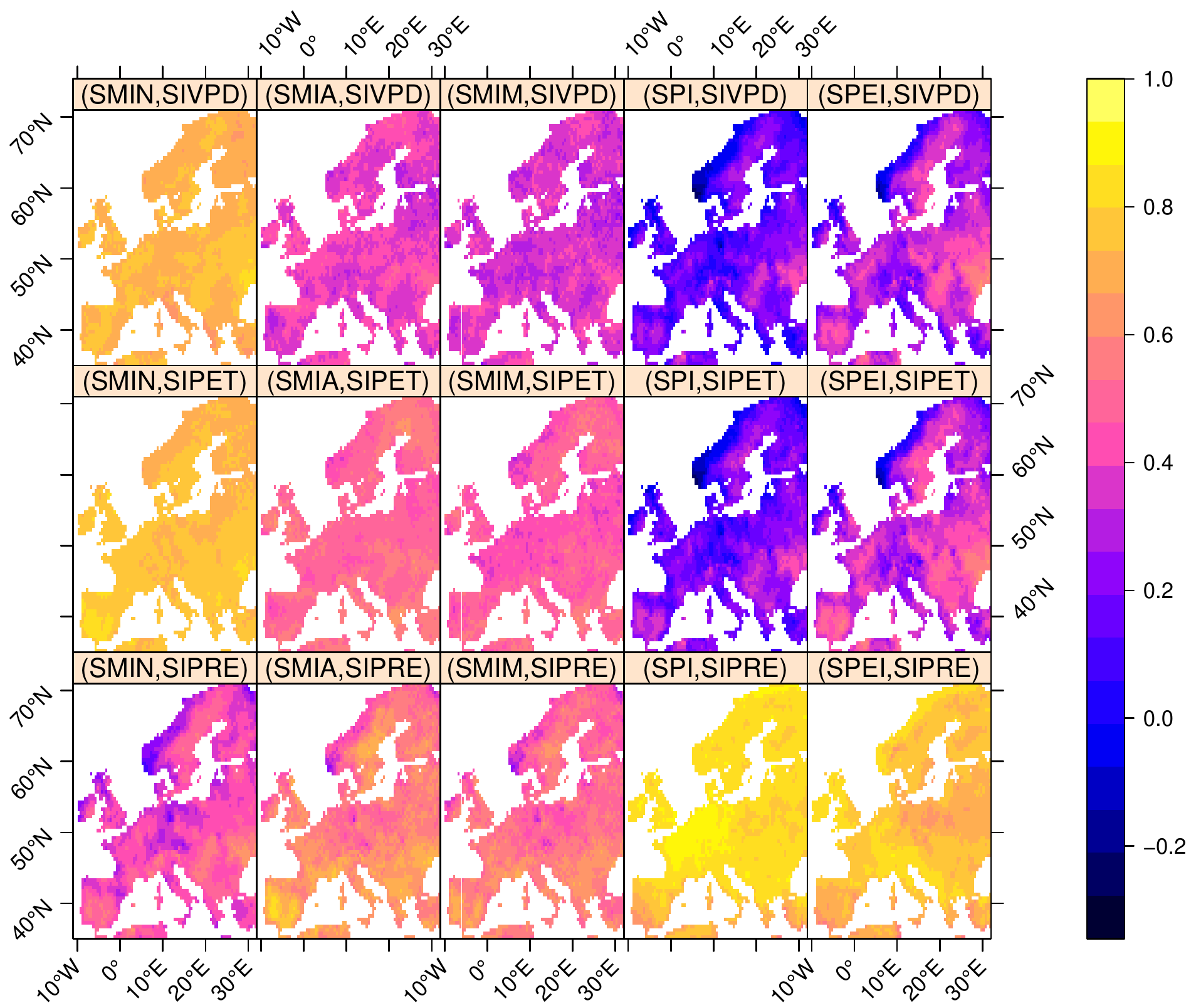}
	\caption{Maps of Kendall's $\tau$ for all combinations of the univariate drought indices $\SI_6(\text{\texttt{VPD}})$ (SIVPD), $\SI_6(\text{\texttt{PET}})$ (SIPET) and $\SI_6(\text{\texttt{PRE}})$ (SIPRE) with the indices $\NSMI_6(\beins;\text{\texttt{VPD}},\text{\texttt{PET}},\text{\texttt{PRE}})$ (SMIN), $\ASMI_6(\beins;\text{\texttt{VPD}},\text{\texttt{PET}},\text{\texttt{PRE}})$ (SMIA), $\MSMI_6(\beins;\text{\texttt{VPD}},\text{\texttt{PET}},\text{\texttt{PRE}})$ (SMIM), $\SPI_6$ (SPI) and $\SPEI_6$ (SPEI).}
	\label{fig:ktau}
\end{figure}

To measure pair-wise dependence we use the rank-based association measure Kendall's $\tau$ \citep[see e.g.][]{kendall70}. In Figure \ref{fig:ktau} we provide maps of Kendall's $\tau$ between the univariate drought indices $\SI_6(\text{\texttt{VPD}})$, $\SI_6(\text{\texttt{PET}})$ and $\SI_6(\text{\texttt{PRE}})$ and the (multivariate) drought indices $\NSMI_6(\beins;\text{\texttt{VPD}},\text{\texttt{PET}},\text{\texttt{PRE}})$, $\ASMI_6(\beins;\text{\texttt{VPD}},\text{\texttt{PET}},\text{\texttt{PRE}})$, $\MSMI_6(\beins;\text{\texttt{VPD}},\text{\texttt{PET}},\text{\texttt{PRE}})$, $\SPI_6$ and $\SPEI_6$ on time scale $6$, to see how the different variables contribute to the different drought indices and how this contribution varies over space. Whereas the $\NSMI$ is dominated by \texttt{PET} and \texttt{VPD} (high Kendall's $\tau$ values all over Europe for the pairs (SMIN,SIVPD) and (SMIN,SIPET)), the other indices are stronger associated with \texttt{PRE} (comparatively high Kendall's $\tau$ values for the pairs (SMIA,SIPRE), (SMIM,SIPRE), (SPI,SIPRE) and (SPEI,SIPRE)). For $\ASMI$ and $\MSMI$ the overall association with \texttt{PET} and \texttt{VPD} is stronger compared to $\SPI$ and $\SPEI$ (compare the corresponding pairs). Especially for $\SPI$ and $\SPEI$ we observe spatial differences in Kendall's $\tau$ (see all pairs involving SPI and SPEI).

To validate and compare the different drought indices we consider the three major drought events of the $30$ years period $1975-2004$ which were observed in large parts of Europe. These droughts occured in the years $1976$, $1989/90$ and $2003$. We summarize these events in Table \ref{tab:drought}. It gives the dates when the drought events (in terms of an extreme ($D3$) or exceptional ($D4$) drought) reached their maximum spatial extent (i.e. the month in which the area affected by a $D3$ or $D4$ drought reached it's maximum) and the corresponding percentage of area under consideration which was affected by an extreme ($D3$) or exceptional ($D4$) drought.

\begin{table}
\caption{\label{tab:drought}Maximum spatial extent of drought events classified as extreme ($D3$) or exceptional ($D4$) according to $\SPI_6$, $\SPEI_6$, $\ASMI_6(\beins;\text{\texttt{VPD}},\text{\texttt{PET}},\text{\texttt{PRE}})$ and $\MSMI_6(\beins;\text{\texttt{VPD}},\text{\texttt{PET}},\text{\texttt{PRE}})$.}
\centering
\fbox{%
\begin{tabular}{l|rr|rr|rr|rr}
& \multicolumn{4}{c}{univariate}  &	\multicolumn{4}{|c}{multivariate}	\\
& \multicolumn{2}{c}{$\SPI_6$}  & \multicolumn{2}{|c}{$\SPEI_6$}  &	\multicolumn{2}{|c}{$\ASMI$}  &	\multicolumn{2}{|c}{$\MSMI$}	\\ 
event	&	max.	&	$\%$ area	&	max.	&	$\%$ area	&	max.	&	$\%$ area	&	max.	&	$\%$ area	\\
\hline
$1976$	&	$07.1976$	&	$31.0\%$	&	$08.1976$	&	$28.4\%$	&	$08.1976$	&	$28.7\%$	&	$08.1976$	&	$24.1\%$	\\
$1990$	&	$03.1990$	&	$18.3\%$	&	$03.1990$	&	$21.3\%$	&	$05.1990$	&	$25.7\%$	&	$05.1990$	&	$36.1\%$	\\
$2003$	&	$08.2003$	&	$21.9\%$	&	$08.2003$	&	$37.2\%$	&	$08.2003$	&	$50.7\%$	&	$08.2003$	&	$46.9\%$
\end{tabular}}
\end{table}

\begin{figure}[!htb]
	\centering
		\includegraphics[width=1.00\textwidth]{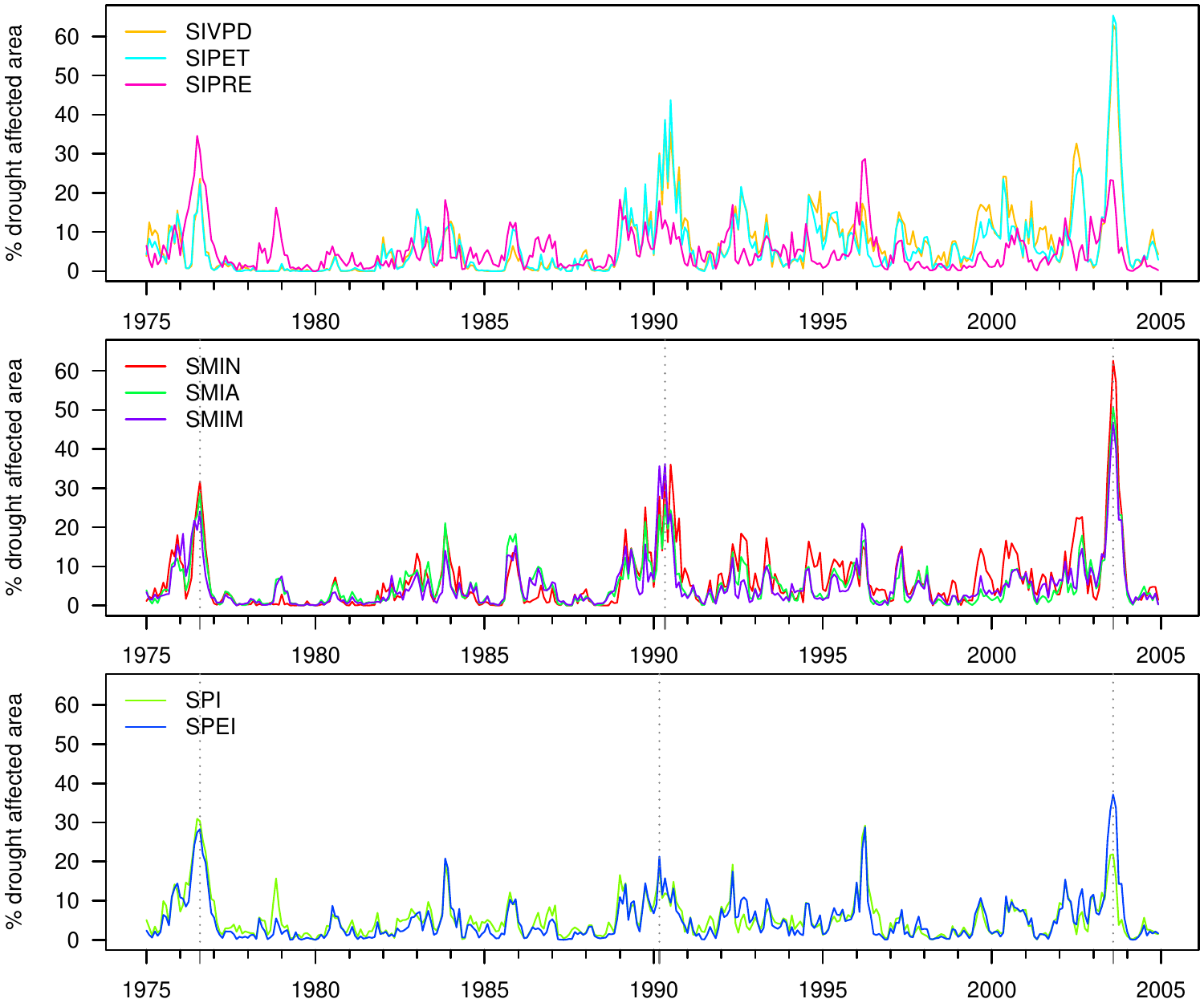}
	\caption{Percentage of area affected by a $D3$ or $D4$ drought according to $\SI_6(\text{\texttt{VPD}})$, $\SI_6(\text{\texttt{PET}})$ and $\SI_6(\text{\texttt{PRE}})$ (upper panel), $\NSMI_6(\beins;\text{\texttt{VPD}},\text{\texttt{PET}},\text{\texttt{PRE}})$, $\ASMI_6(\beins;\text{\texttt{VPD}},\text{\texttt{PET}},\text{\texttt{PRE}})$ and $\MSMI_6(\beins;\text{\texttt{VPD}},\text{\texttt{PET}},\text{\texttt{PRE}})$ (middle panel), and $\SPI_6$ and $\SPEI_6$ (lower panel).}
	\label{fig:DroughtArea}
\end{figure}

Figure \ref{fig:DroughtArea} compares time series of the percentage of area affected by drought according to the different univariate and multivariate drought indices calculated following the methodology described above, as well as $\SPI_6$ and $\SPEI_6$. Comparing the univariate indices we see that those based on \texttt{PET} and \texttt{VPD} yield similar however not identical results. During the three major drought events in $1976$, $1989/90$ and $2003$ all three univariate indices indicate extreme dry conditions for large parts of Europe. Comparison to the middle panel shows that the multivariate indices successfully combine the drought information captured in the single variables used for their calculation. For the years $1990$ and $2003$ abnormally high \texttt{PET} and \texttt{VPD} aggravate the dry conditions due to a lack of precipitation. During the years $1994$, $1995$, $1999$ and $2000$ one can see that the vine copula based indices are more conservative compared to $\NSMI$, since they are not as much influenced by \texttt{PET} and \texttt{VPD}. In terms of spatial extent the multivariate indices classify the drought events of $1990$ and $2003$ as more severe compared to $\SPI$ and $\SPEI$.

%% file: Sections/Conclusions.tex
\section[Conclusions]{Conclusions and outlook}\label{sec:conc}

\begin{figure}[!htb]
	\centering
\begin{tikzpicture}[node distance=4cm]
\node (start) [start] {Input: $d$ time series of drought relevant variables (\texttt{ARBVAR})};
\node (pro1) [process, right of=start] {1. Variable transformation (skewness reduction, \texttt{DRYWET})};
\node (pro2) [process, right of=pro1] {2. Elimination of seasonality (\texttt{SEASON}, \texttt{TRENDS}, \texttt{SMALLS})};
\node (pro3) [process, right of=pro2] {3. Elimination of serial dependence (select AR-/MA-order, \texttt{TIMDEP})};
\node (pro4) [process, below of=pro3, yshift=+1cm] {4. Marginal transformation (PIT) to u-scale/ copula data (\texttt{NPDIST})};
\node (pro5) [process, left of=pro4] {5. Dependency modeling (vine copula selection, Rosenblatt transform, \texttt{MULTEX})};
\node (dec1) [decision, left of=pro5] {Decision on: method ($\calN$, $\calA$, $\calM$), weights $\bw$, time scale $l$ (\texttt{TSCALE})};
\node (stop1) [stop, below of=dec1, yshift=1.25cm] {$\ASMI_l(\bw;1,\ldots,d)$};
\node (stop2) [stop, left of=stop1] {$\NSMI_l(\bw;1,\ldots,d)$};
\node (stop3) [stop, right of=stop1] {$\MSMI_l(\beins;1,\ldots,d)$};
\node (stop4) [stop, right of=stop3] {(\texttt{STCOMP})};
\draw [arrow] (start) -- (pro1);
\draw [arrow] (pro1) -- (pro2);
\draw [arrow] (pro2) -- (pro3);
\draw [arrow] (pro3) -- (pro4);
\draw [arrow] (pro4) -- (pro5);
\draw [arrow] (pro5) -- (dec1);
\draw [arrow] (dec1) -- node[anchor=east] {$\calA$} (stop1);
\draw [arrow] (dec1) -- node[anchor=east] {$\calN$} (stop2);
\draw [arrow] (dec1) -- node[anchor=east] {$\calM$} (stop3);
\end{tikzpicture}
\caption{Modeling steps for multivariate drought index calculation.}
	\label{fig:fchart}
\end{figure}
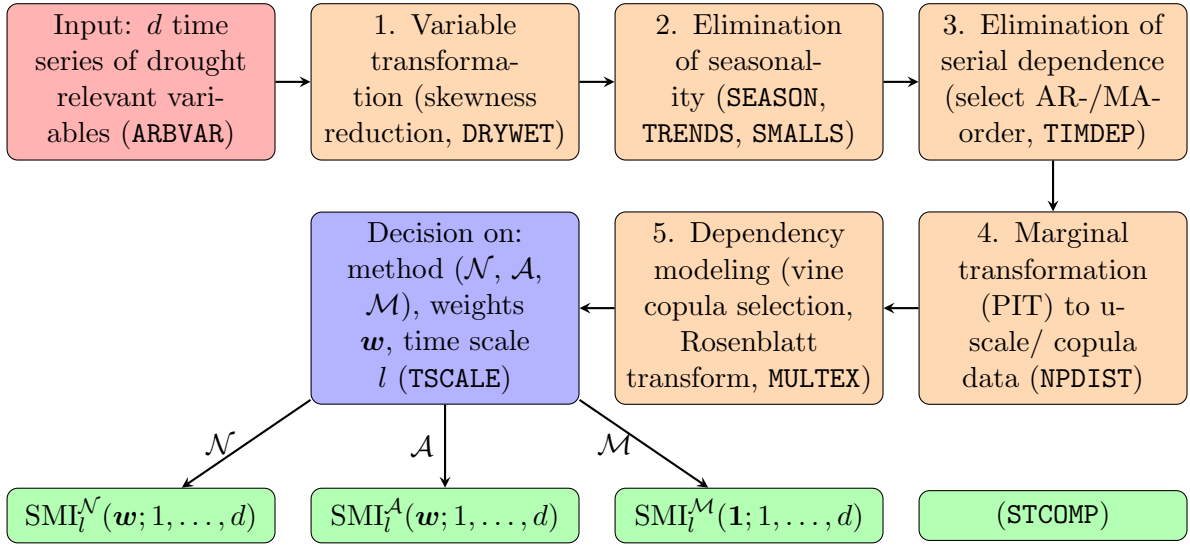

Comparison of the advantages and disadvantages of existing drought indices and the flexibility of vine copulas in modeling multivariate dependence structures led to a novel and flexibly applicable approach to calculate drought indices based on arbitrary sets of drought relevant variables. This approach involves several well reasoned modeling steps which we summarize in Figure \ref{fig:fchart}.

Taking several drought drivers and their dependencies into account at the same time our novel approach enables flexible modeling of different drought types and allows tailoring of drought indices to specific applications. An example would be the application of the novel methodology in the field of ecology. Multivariate drought indices based on selected variables could be calibrated to tree ring data to find good models for the response of tree growth to climatic conditions. Moreover, the presented approach for the calculation of severity indices is not restricted to drought. Applications to model for example the degree of contamination of a water body due to different contaminants are feasible.